\documentclass{PoS}

\title{Precision Measurement of $\sin^2 \theta_w$ at MESA 
       \thanks{Supported by the Collaborative Research Center 
               1044 funded through the Deutsche 
               Forschungsgemeinschaft (DFG).}
}

\ShortTitle{Precision Measurement of $\sin^2 \theta_w$ at MESA}

\author{R Bucoveanu\\
        PRISMA Cluster of Excellence, Institut f\"ur Physik,
        Johannes Gutenberg-Universit\"at, 55099 Mainz, Germany\\
        E-mail: \email{rabucove@uni-mainz.de}}

\author{M Gorchtein\\
        PRISMA Cluster of Excellence, Institut f\"ur Kernphysik,
        Johannes Gutenberg-Universit\"at, 55099 Mainz, Germany\\
        E-mail: \email{gorshtey@kph.uni-mainz.de}}

\author{\speaker{H Spiesberger}
        \\
        PRISMA Cluster of Excellence, Institut f\"ur Physik,
        Johannes Gutenberg-Universit\"at, 55099 Mainz, Germany,
        and Centre for Theoretical and Mathematical Physics and 
        Department of Physics, University of Cape Town, Rondebosch 
        7700, South Africa\\
        E-mail: \email{spiesber@uni-mainz.de}}

\abstract{A forthcoming experiment of low-energy elastic 
        electron proton scattering at the new MESA facility 
        in Mainz is planned to provide a high-precision measurement 
        of the parity-violating polarisation asymmetry. This 
        experiment is expected to lead to a precision determination 
        of the weak mixing angle, competitive with $Z$-pole data. 
        We discuss the challenges for theory to derive predictions 
        with the required accuracy.}

\FullConference{Loops and Legs in Quantum Field Theory\\
		24-29 April 2016\\
		Leipzig, Germany}

\begin{document}

\section{The running weak mixing angle}

\begin{figure}
\begin{center}
\includegraphics[width=.7\textwidth]{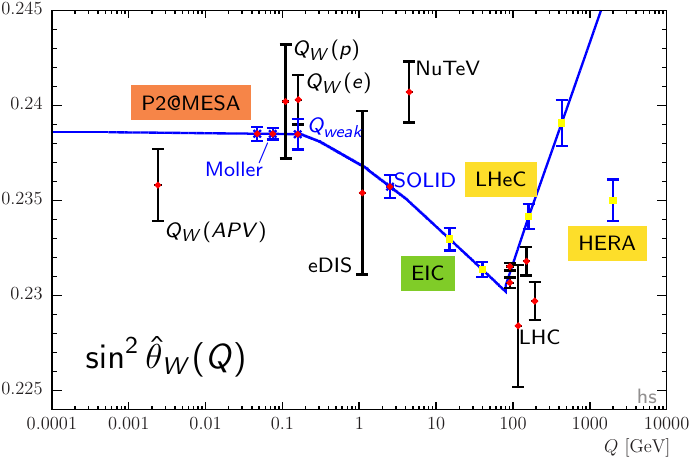}
\end{center}
\caption{Running weak mixing angle compared with data. 
Existing measurements are shown in black for atomic parity 
violation in Cesium ($Q_W(APV)$), parity-violating electron 
scattering off protons ($Q_W(p)$ and eDIS), and electrons 
($Q_W(e)$), neutrino nucleon scattering (NuTeV). Data points 
at the $Z$ pole are from LEP1, SLD, Tevatron, CMS and ATLAS. 
Blue symbols show results of possible future experiments: 
P2@MESA, the Moller, Qweak and SOLID experiments at JLAB, an 
electron-ion collider (EIC), the LHeC and an estimate for the 
analysis of existing data from the HERA experiments. See also 
Ref.\ \cite{Agashe:2014kda}. The scale dependence of the weak 
mixing angle was calculated using the program described in 
\cite{Erler:1999ug}.}
\label{HS:fig1}
\end{figure}

One of the central parameters in the theory of the electroweak 
interaction is the weak mixing angle. In the Standard Model 
(SM) it can be defined as a scale-dependent quantity. Many 
measurements have confirmed the SM prediction for its running, 
as shown in Fig.\ \ref{HS:fig1}. The most precise single 
measurements at the $Z$ pole from LEP1 and SLD are only 
marginally consistent with each other and additional data 
with similarly high accuracy are required to improve the 
strength of SM tests as well as limits from searches for new 
physics. 

\section{The P2 experiment at MESA}

The new accelerator MESA (Mainz Energy-Recovery Superconducting 
Accelerator) being built at Mainz University will provide an 
electron beam with an energy up to $E=155$ MeV and with a high 
degree of longitudinal polarization (above 85\, \%). The P2 
experiment will measure the parity-violating asymmetry between 
left- and right-handed electrons, 
\begin{equation}
 A_{LR} 
 = 
 \frac{\sigma(e_\downarrow) - \sigma(e_\uparrow)}
      {\sigma(e_\downarrow) + \sigma(e_\uparrow)}
 = 
 - \frac{G_F Q^2}{4\sqrt{2}\pi \alpha} 
 \left(Q_W(p) - F(Q^2)\right)
\end{equation}
at scattering angles in the range from 25 to 45 degrees, 
corresponding to an averaged squared momentum transfer $Q^2$ 
of about 0.0045 GeV$^{2}$. $A_{LR}$ is determined by the weak 
charge of the proton, given in the SM at leading order by the 
weak mixing angle $\sin^2 \theta_W$, 
\begin{equation}
Q_W(p) = 1 - 4 \sin^2 \theta_W \, .
\end{equation}
The experimental set-up aims to measure $A_{LR}$ with a 
total uncertainty of 1.5\, \% which, by error propagation, 
leads to a 0.13\, \% measurement of $\sin^2 \theta_W$. 

At non-zero $Q^2$, $A_{LR}$ is affected by form factor 
contributions $F(Q^2)$ due to the fact that the proton is 
not a point-like particle. $F(Q^2)$ and consequently $A_{LR}$ 
can be decomposed as \cite{Musolf:1993tb}
\begin{eqnarray}
 F(Q^2) 
 &=& 
 F_{\mathrm{EMFF}}(Q^2) 
 + F_{\mathrm{Axial}}(Q^2) 
 + F_{\mathrm{Strangeness}}(Q^2) \, , 
\nonumber \\
 A_{LR} 
 &=&
 A_{\mathrm{Q_{W}}} 
 + A_{\mathrm{EMFF}} 
 + A_{\mathrm{Axial}} 
 + A_{\mathrm{Strangeness}} 
\end{eqnarray}
into contributions determined by electric and magnetic form 
factors of the proton and neutron, $G_E^{p,n}$, $G_M^{p,n}$
\begin{equation}
 F_{\mathrm{EMFF}}(Q^2) 
 = 
 \frac{\epsilon G_E^p G_E^n 
 + \tau G_M^p G_M^n}{\epsilon (G_E^p)^2 + \tau (G_M^p)^2} \, ,
\end{equation}
the axial proton form factor $G_A^p$
\begin{equation}
 F_{\mathrm{Axial}}(Q^2) 
 = 
 \frac{(1-4 \sin^2\theta_W)
       \sqrt{1-\epsilon^2}\sqrt{\tau (1+\tau)} G_M^p G_A^p} 
 {\epsilon (G_E^p)^2 + \tau (G_M^p)^2} \, ,
\end{equation}
and a part containing the strangeness form factors $G_E^s$, 
$G_M^s$
\begin{equation}
 F_{\mathrm{Strangeness}}(Q^2) 
 = 
 \frac{\epsilon G_E^p G_E^s 
 + \tau G_M^p G_M^s}{\epsilon (G_E^p)^2 + \tau (G_M^p)^2}
 \, . 
\end{equation}
Here we have used the usual kinematic variables 
\begin{equation}
 \epsilon = [1 + 2(1 + \tau) \tan^2(\theta/2) ]^{-1} \, ,
 \qquad 
 \tau = Q^2/4 m_p^2 
\end{equation}
and $m_p$ is the proton mass.

\begin{figure}[b]
\begin{center}
\includegraphics[width=.48\textwidth]{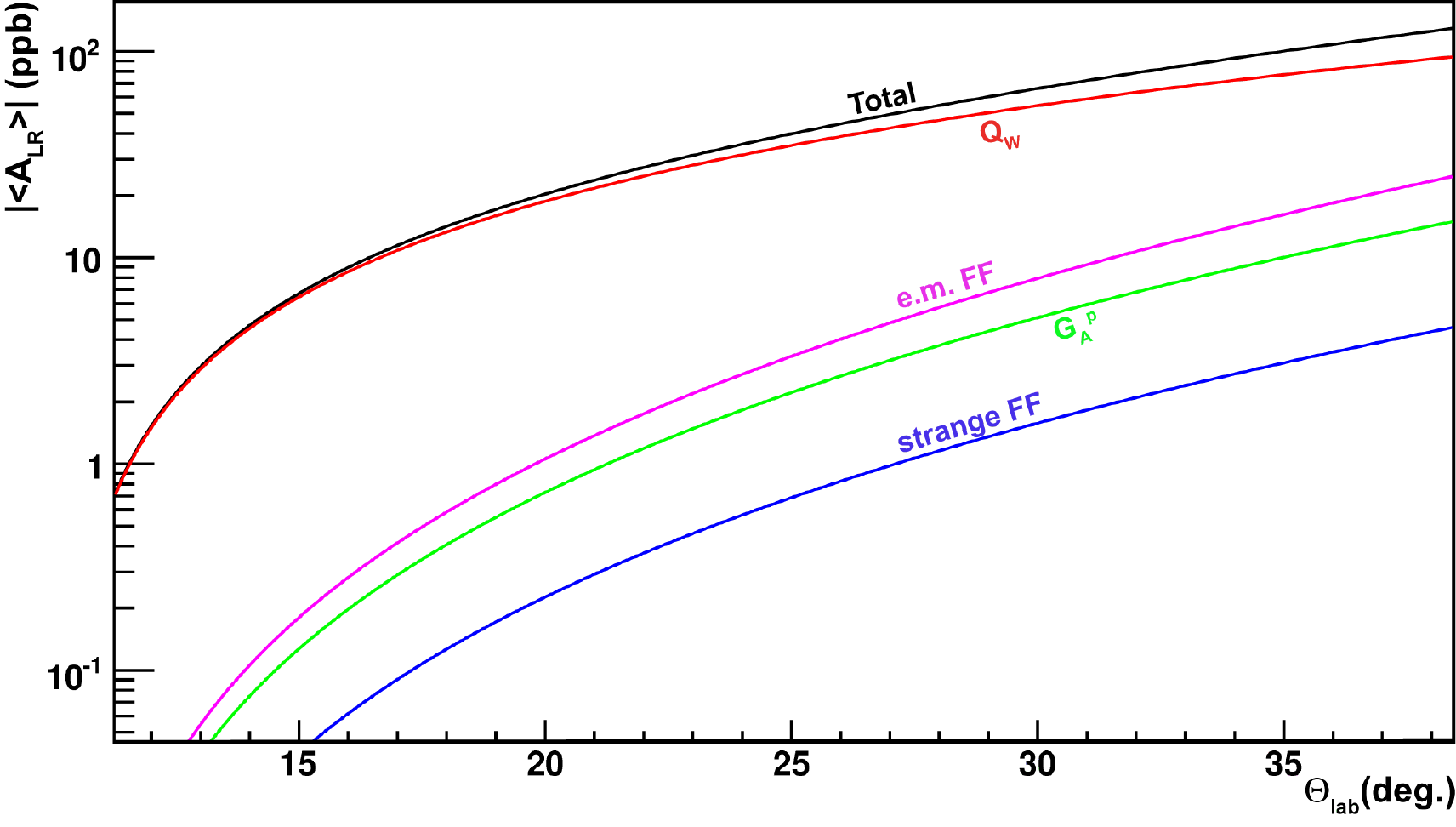}
\includegraphics[width=.48\textwidth]{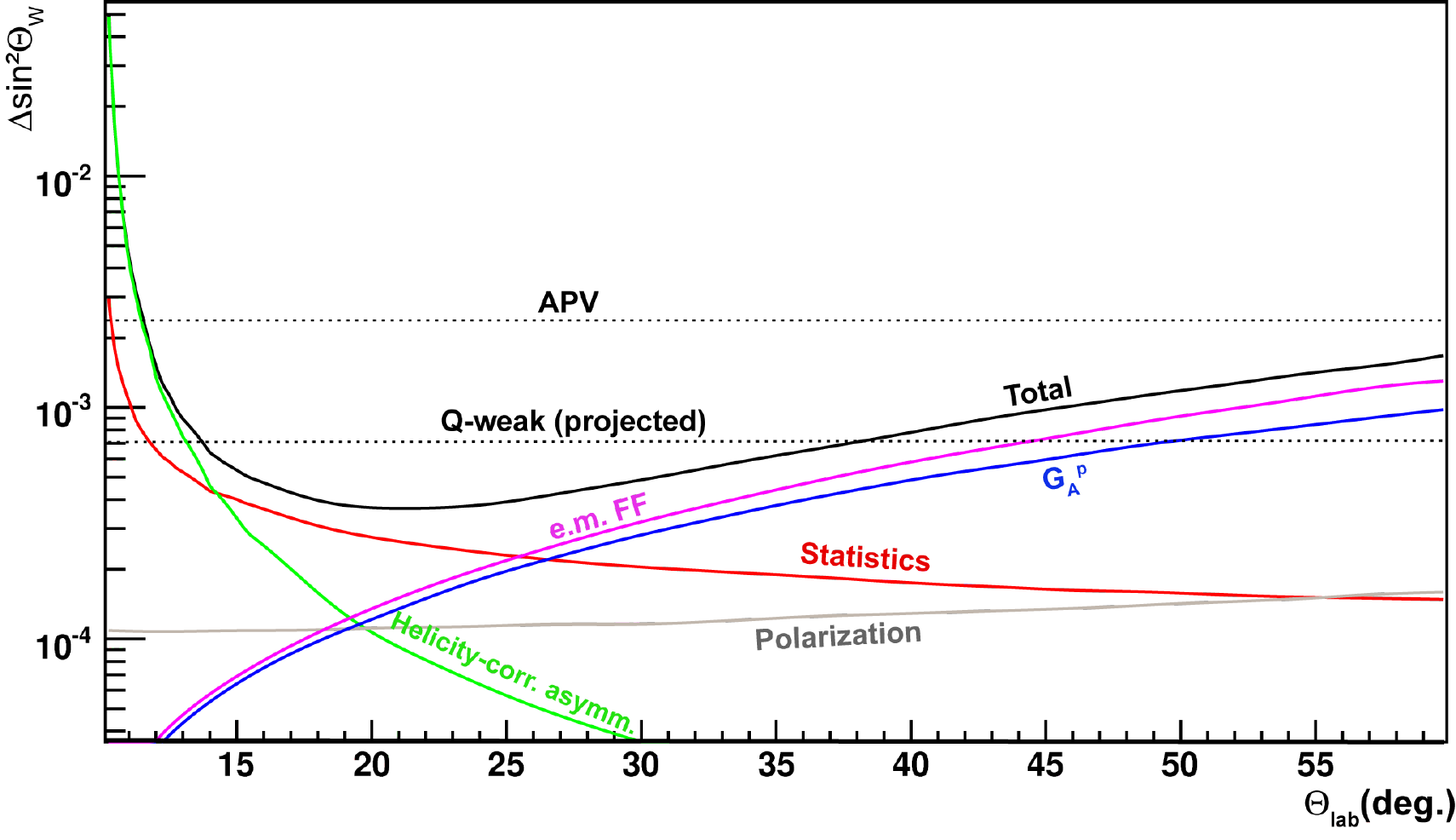}
\end{center}
\caption{Left: proton weak charge and form factor contributions 
to $A_{LR}$ as a function of the electron scattering angle for 
the beam energy $E=200$ MeV. Right: estimated uncertainties 
for $\sin^2 \theta_W$ at $E=200$ MeV due to form factors, 
statistics, polarization measurement and helicity-correlated 
beam asymmetries. Figures taken from \cite{Becker:2013cma, 
Becker:2014hta}.}
\label{HS:fig2}
\end{figure}

As an illustration we show the asymmetry and the uncertainty 
for the weak mixing angle as a function of the scattering 
angle at a beam energy of $E=200$ MeV in Fig.\ \ref{HS:fig2} 
\cite{Becker:2013cma,Becker:2014hta}. Results for lower 
energies are similar. From the left panel of Fig.\ \ref{HS:fig2} 
we can see that $A_{LR}$ is indeed dominated by the proton 
weak charge; form factor contributions are suppressed by about 
an order of magnitude in the range of scatterings angles 
relevant for the P2 experiment. In the right panel of Fig.\ 
\ref{HS:fig2}, a break-down of the uncertainties for $\sin^2 
\theta_W$ is shown. At low $Q^2$, the statistical uncertainty 
and uncertainties due to helicity-correlated beam fluctuations 
are dominating. Form factor uncertainties are increasing 
with the scattering angle, but a minimum of the total 
uncertainty can be found at scattering angles corresponding 
to $Q^2 = 0.0045$ GeV$^{2}$. Details for the accelerator 
and detector systems and the polarimetry are described in 
\cite{Aulenbacher:2013xla,Berger:2015aaa,Aulenbacher:2012um}. 

\section{Theory challenges}

The prediction for the left-right asymmetry is affected by 
higher-order corrections, 
\begin{equation}
 A_{LR} = 
 - \frac{G_F Q^2}{4\sqrt{2}\pi \alpha}  
 \left(Q_W(p) (1 + \delta_1) - \widetilde{F}(Q^2)\right)
\end{equation}
where $\delta_1$ comprises universal correction factors 
$\rho_{\rm NC}$ and $\kappa$ as well as process-specific 
corrections due to vertex ($\Delta_e$, $\Delta_e^\prime$) 
and box graph contributions ($\delta_{\rm Box}$), 
\begin{equation}
 Q_W(p)(1 + \delta_1) 
 = \left(\rho_{\rm NC} + \Delta_e\right) 
 (1 - 4 \kappa \sin^2 \theta_W 
 + \Delta_e^\prime) + \delta_{\rm Box} \, .
\end{equation}
In particular, photon-$Z$ boson mixing contributes to 
corrections that can be absorbed into a scale-dependent 
weak mixing angle, 
\begin{equation}
 \sin^2 \theta_{\rm eff}(\mu^2)
 =
 \kappa(\mu^2) \sin^2 \theta_W \, .
\end{equation}
Depending on the renormalization scheme, $\kappa$ can also 
contain some non-universal loop corrections. At one-loop 
order, these corrections are well-known to a high precision 
\cite{Erler:2003yk,Erler:2004in}. The high accuracy aimed 
for at P2@MESA will, however, require to evaluate also 
two-loop corrections. For M{\o}ller scattering, first steps 
towards a complete two-loop calculation have been made 
\cite{Aleksejevs:2013gxa,Aleksejevs:2015dba,Aleksejevs:2015zya} 
and show that their effect on the measured asymmetry may 
be larger than naively expected. 

\subsection{QED corrections}

\begin{figure}[t]
\begin{center}
\includegraphics[width=.6\textwidth]{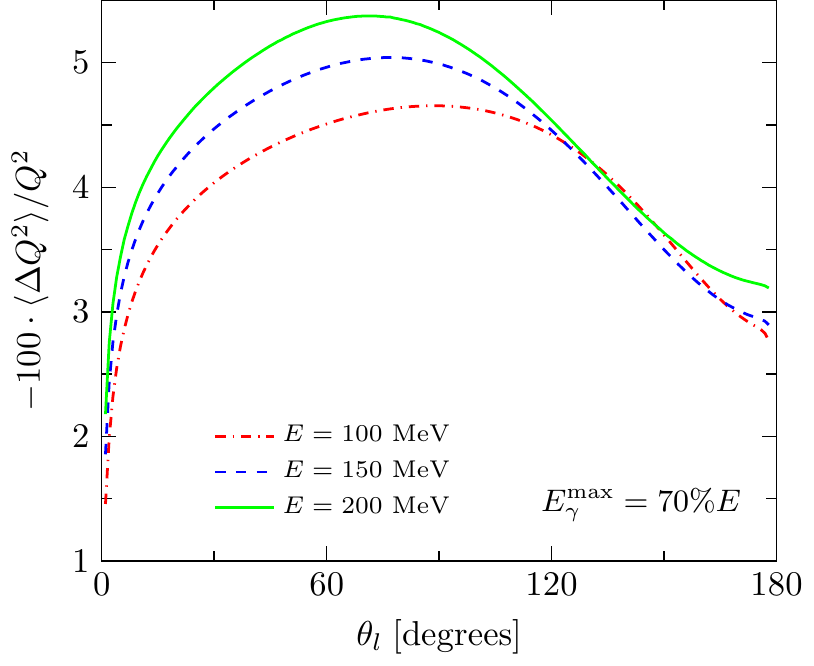}
\end{center}
\caption{Shift of the momentum transfer due to photon 
bremsstrahlung.}
\label{HS:fig3}
\end{figure}

Electromagnetic corrections are parity-conserving and do 
not affect $A_{LR}$ directly. However, the momentum transfer 
has to be known with high precision in order to extract the 
weak charge from the measured asymmetry. Therefore, 
bremsstrahlung effects have to be calculated as well since 
they lead to a shift of the momentum transfer measured from 
the scattering angle of the electron relative to the true 
momentum transferred to the proton. In Fig.\ \ref{HS:fig3} 
we show the results of a calculation including one-photon 
bremsstrahlung. The $Q^2$-shift depends strongly on the 
beam energy and the scattering angle, as well as on a 
possible cutoff of the energy of photons radiated into the 
final state. It is obvious from these results that also 
the dominating two-photon bremsstrahlung contributions 
will have to be evaluated in order to reach the high-precision 
goal of P2@MESA.

\subsection{$\gamma Z$ box graphs}

An important source of uncertainties in higher-order corrections 
is due to box graph contributions with the exchange of a photon 
and a $Z$ boson. The presence of the massless photon in the loop 
makes these graphs sensitive to the low-scale hadronic structure 
of the proton. 

\begin{figure}[b]
\begin{center}
\includegraphics[width=.6\textwidth]{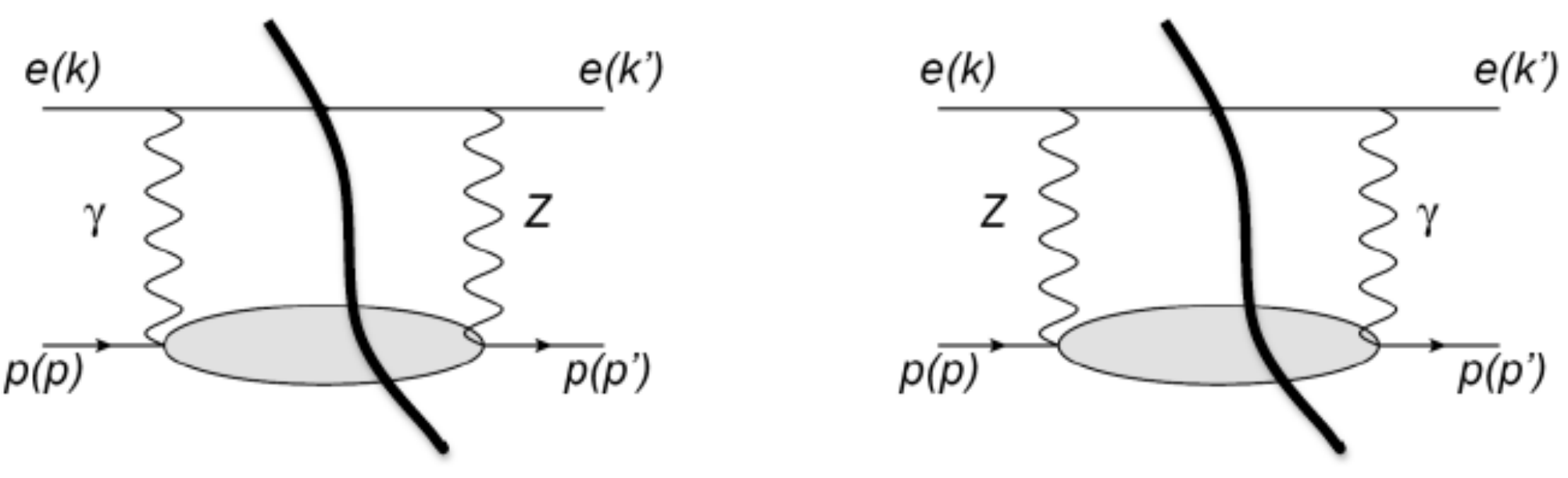}
\end{center}
\caption{The optical theorem and dispersion relations 
allow one to relate the box graph corrections to elastic 
electron proton scattering with structure functions for 
inelastic scattering.}
\label{HS:fig5}
\end{figure}

The optical theorem can be used to relate the imaginary part 
of the box graphs to the $\gamma Z$ interference part of 
structure functions $F_k^{\gamma Z}$ for inelastic $ep$ 
scattering (see Fig.\ \ref{HS:fig5}), 
\begin{equation}
 {\rm Im} \Box_{\gamma Z}(E) 
 = 
 \frac{\alpha}{(s-m_p^2)^2} 
 \int_{W_\pi^2}^s dW^2 
 \int_0^{Q^2_{max}} dQ^2 \frac{M_Z^2}{Q^2+M_Z^2}
 \left\{
 F_1^{\gamma Z} + A F_2^{\gamma Z}
 + \frac{g_V^e}{g_A^e} BF_3^{\gamma Z}
 \right\} 
\end{equation}
where $s = 2m_pE+m_p^2$ is the squared center-of-mass energy, 
$W$ the invariant mass of the hadronic intermediate state, 
$\nu$ the invariant energy transfer, $W^2 = m_p^2 + 2m_p \nu 
- Q^2$, $A$ and $B$ are kinematic factors, $g_V^e$, $g_A^e$ 
the weak neutral-current vector and axial-vector coupling 
constants of the electron and the $F_k^{\gamma Z}$ are 
functions of $\nu$ and $Q^2$. The dispersion relations 
involve integrals over the full kinematic range, with a 
strong emphasis on the low-$W$, low-$Q^2$ range. The structure 
functions can, in principle, be measured in $ep$ scattering. 
In practice, however, data are available only in a very 
restricted range of the kinematic variables. Missing 
information has to be modelled, for example by assuming 
dominance of low-lying resonances in the hadronic 
intermediate state of the box graph, or in baryon chiral 
perturbation theory. 

\begin{figure}
\begin{center}
\includegraphics[width=.6\textwidth]{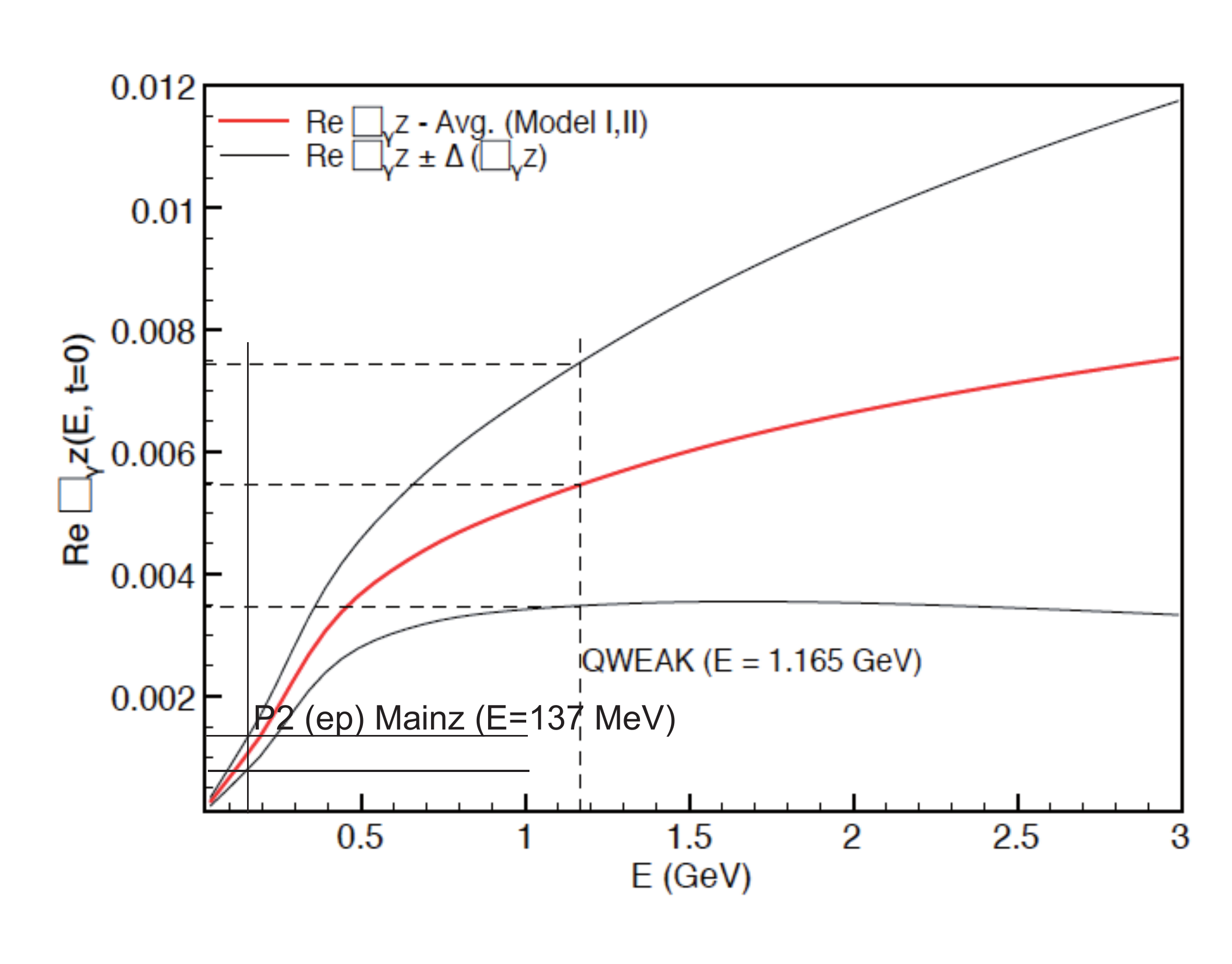}
\end{center}
\caption{$\gamma Z$ box graph correction with its error 
estimate from Ref.\ \cite{Gorchtein:2011mz}. The full and 
dashed lines indicate the beam energies and resulting 
uncertainties for the P2@MESA and Qweak experiments.}
\label{HS:fig6}
\end{figure}

The box graph contributions can be separated into vector 
and axial-vector parts of the proton current and the 
real parts that enter the corrections for $A_{LR}$ are 
recovered by dispersion relations which read  
\begin{eqnarray}
 {\rm Re} \Box_{\gamma Z}^V(E) 
 & = & 
 E \frac{2}{\pi} 
 \int_{\nu_\pi}^{\infty} \frac{dE^\prime}{E^{\prime\, 2}-E^2}
 {\rm Im} \Box_{\gamma Z}^V(E^\prime) \, ,
 \label{gzboxv}
 \\
 {\rm Re} \Box_{\gamma Z}^A(E) 
 & = & 
 \frac{2}{\pi} 
 \int_{\nu_\pi}^{\infty} 
 \frac{E^\prime dE^\prime}{E^{\prime\, 2} - E^2}
 {\rm Im} \Box_{\gamma Z}^A(E^\prime) \, . 
\end{eqnarray}
Invariance with respect to time reversal forces the 
vector part Eq.\ (\ref{gzboxv}) to vanish for zero beam 
energy. This fact makes the measurement at P2@MESA ($E \leq 
155$ MeV) much less sensitive to theoretical uncertainties 
than the competing experiment Qweak at the Jefferson 
Laboratory ($E = 1.165$ GeV, $Q^2 = 0.026$ GeV$^2$) 
\cite{Androic:2013rhu}. The energy-dependence of the 
vector $\gamma Z$ box graph correction from Ref.\ 
\cite{Gorchtein:2011mz} is shown in Fig.\ \ref{HS:fig6}. 

A recent update \cite{Gorchtein:2015naa} has taken into 
account contributions from strange form factor contributions 
modelled in a unitarized partial wave analysis supplemented 
by a Regge theory inspired high-energy behaviour of the 
structure function input. The currently best estimate for 
the vector part of the $\gamma Z$ box graph correction is 
\begin{equation}
 {\rm Re}\,\Box^{V}_{\gamma Z} 
 (E=0.155\,{\rm GeV}) 
 = 
 (1.07\pm0.18)\times10^{-3} \, ,
\end{equation}  
to be compared with 
\begin{equation}
 {\rm Re}\,\Box^{V}_{\gamma Z}
 (E=1.165\,{\rm GeV}) 
 =  
 (5.58\pm1.41)\times10^{-3} 
\end{equation}  
for the kinematics of the Qweak experiment. This numerical 
result, a 0.1\,\% correction for $A_{LR}$ for P2@MESA, 
is well below the required accuracy and confirms the 
expectation that a measurement at lower beam energy is 
advantageous. 

Parity violation in the hadronic system will also contribute 
through $\gamma \gamma$ box diagrams. Their study is 
presently underway \cite{MGHS16}. 

The conventional formal definition of the nucleon's weak charge 
through the low-momentum limit of a measured parity-violating 
asymmetry is based on the assumption that form factor 
contributions vanish at $Q^2 = 0$ and can be written as 
$F(Q^2) = - Q^2 B(Q^2)$. With this assumption one could 
write
\begin{equation}
A_{exp} = A_0
\left(Q_W(p) + Q^2 B(Q^2)\right)
\quad \rightarrow \quad
Q_W(p) = \lim_{Q^2 \rightarrow 0} \frac{A_{exp}}{A_0}
\, .
\end{equation}
However, this definition ignores the presence of box graph 
corrections which depend on both $Q^2$ and the beam 
energy, but do not vanish at $Q^2$. A revised definition 
of the weak charge, taking this observation into account, 
is 
\begin{equation}
A_{exp} = A_0
\left(Q_W(p) + Q^2 B(Q^2) + \Box(E) \right)
\quad \rightarrow \quad
Q_W(p) = \lim_{Q^2, E \rightarrow 0} \frac{A_{exp}}{A_0}
\end{equation}
where the zero-scale limit of both the momentum transfer 
and the center-of-mass energy has to be taken.

\section{Conclusions}

Civil construction for the extremely challenging high-precision 
experiments at MESA will start 2016 and we expect that a beam 
for the P2 experiment will be available before 2020. In the 
larger context of past, present and future $ep$ scattering 
experiments, MESA will be the facility with the highest beam 
intensity at lowest electron energy, reaching an integrated 
luminosity of almost 10\,ab$^{-1}$ in 10,000 hours of data 
taking. The P2 experiment will measure the smallest 
PV-violating asymmetry, $A_{LR} \simeq 33$ ppb with highest 
precision $\Delta A_{LR} \simeq 0.44$ ppb. 

The expected precision for $\sin^2 \theta_W$ from this 
experiment will be competitive with the highest-precision 
results from $Z$-pole data, both from LEP1 and from expected 
LHC measurements. In searches for new physics, the precision 
will allow us to exclude 4-fermion contact interactions 
with mass scales up to 49 TeV, corresponding to the reach 
at the LHC with an integrated luminosity of 300 fb$^{-1}$. 
The planned experiment will be able to exclude models which 
change the running of the weak mixing angle, induced for 
example by new light gauge bosons \cite{Davoudiasl:2014kua}, 
thus complementing other searches for so-called dark photons 
or $Z$-bosons.


\end{document}